\documentclass[10pt,twocolumn,prl,aps,amssymb,amsmath,tightenlines,showpacs]{revtex4}
\usepackage{dsfont}
\usepackage[dvips]{graphicx}
\usepackage{amssymb,amsfonts, color}
\usepackage{overpic}

\newcommand{\re}{\mbox{$\rm e$}}
\newcommand{\ri}{\mbox{$\rm i$}}
\newcommand{\rd}{\mbox{$\rm d$}}

\begin{document}

\title{Solution to the quantum Zermelo navigation problem}

\author{Dorje C. Brody$^{1,2}$ and David M. Meier$^1$}

\affiliation{$^1$Department of Mathematics, Brunel University, Uxbridge UB8 3PH, UK  \\ 
              $^2$St Petersburg State University of Information Technologies, Mechanics and Optics, \\ 
              Kronwerkskii ave 49, St Petersburg 197101, Russia }

\date{September 11, 2014}

\begin{abstract}
The solution to the problem of finding a time-optimal control Hamiltonian to generate a given unitary gate, in an environment in which there exists an uncontrollable ambient Hamiltonian (e.g., a background field), is obtained. In the classical context, finding the time-optimal way to steer a ship in the presence of a background wind or current is known as the Zermelo navigation problem, whose solution can be obtained by working out geodesic curves on a space equipped with a Randers metric. The solution to the quantum Zermelo problem, which is shown here to take a remarkably simple form, is likewise obtained by finding explicit solutions to the geodesic equations of motion associated with a Randers metric on the space of unitary operators. The result reveals that the optimal control in a sense `goes along with the wind'. 
\end{abstract}


\maketitle




The problem of finding the optimal Hamiltonian for processing a given quantum state, or 
implementing a quantum operation (gate), in shortest possible time subject to certain 
constraints, has attracted  considerable attention over the past decade 
\cite{brockett,schulte,zanardi,caneva,bloch,lee,GCH,garon}. Broadly speaking, 
the task can be classified into two closely-related categories: (a) transforming one quantum 
state into another; and (b) transforming one unitary operator into another, in the shortest 
possible time. If the constraint is concerned merely with a limit on energy resource, then the 
optimal Hamiltonian is time independent, and can be found easily by noting that under a 
unitary motion, the shortest path coincides with the path along which the speed of evolution is also 
maximised \cite{brody1,brody2}. If there are further constraints, for example, the choice of 
the Hamiltonian is limited, then often a time-dependent Hamiltonian that minimises an action 
has to be determined by variational approaches \cite{hosoya,hosoya2}. Finding a solution to such a 
variational problem is in general difficult, however, an efficient numerical regularisation 
scheme to obtain an approximate solution has been proposed more recently \cite{lloyd}. 

For many problems related to controlling quantum systems considered in the literature, it is assumed 
that the experimentalist has  full control over the allowable range of Hamiltonians within the 
constraint, whereas in a laboratory there can often be situations in which the system is 
immersed in an external field or potential that is beyond control (e.g., gravitational or 
electro-magnetic field), since a complete elimination of external fields in a laboratory can be 
prohibitively expensive for the given task. Evidently, this is a generic issue that needs to be 
addressed adequately to be able to accurately implement a rapid quantum processing. 

In the present paper we address this issue by finding the time-optimal control Hamiltonian 
${\hat H}(t)={\hat H}_0+{\hat H}_1(t)$ that generates a unitary motion to transform one unitary 
operator ${\hat U}_I$ into another operator ${\hat U}_F$, subject to the constraints (i) that the 
background Hamiltonian ${\hat H}_0$ cannot be controlled; (ii) that the control Hamiltonian 
fulfils the energy resource bound of the form ${\rm tr}(H_1^2)=1$ at all time; and (iii) that 
the background Hamiltonian is not dominating in the sense that ${\rm tr}({\hat H}_0^2)<
{\rm tr}({\hat H}_1^2)$. This is the quantum counterpart of a well-known classical navigation 
problem posed by Zermelo: given the present location in the ocean, with a given wind and/or 
current distribution characterised by a location-dependent vector field, one wishes to find 
the optimal control of the vessel so as to reach the destination in the shortest possible time 
\cite{zermelo,caratheodory}. The vector field generated by the reference Hamiltonian 
${\hat H}_0$ can be thought of as representing the background wind or current, whereas 
${\hat H}_1$ determines the control. 

In the classical context, it was observed by Shen \cite{shen} that the solution to the Zermelo 
navigation problem can be obtained by finding the geodesic curves associated with a Randers 
metric on the configuration space. Motivated by this result, more recently Russell \& Stepney 
\cite{stepney} introduced the quantum Zermelo navigation problem stated above, and analysed 
the shortest time required to realise the transformation ${\hat U}_I\to{\hat U}_F$. Their 
observation that quantum Zermelo navigation problems can be solved by finding the geodesics 
of Randers metrics opens up the possibility of addressing a wide range of realistic quantum 
control problems where the environmental influence cannot be eliminated. However, analyses 
involving Randers spaces are generally difficult, and finding solutions to the geodesic 
equations is not straightforward \cite{Yasuda,BCS}. Indeed, the only examples considered in 
\cite{stepney} concern the time-independent cases, while the optimal navigation is realised by a 
time-independent Hamiltonian only if the background Hamiltonian ${\hat H}_0$ happens to be the one that 
realises the operation ${\hat U}_I\to{\hat U}_F$. But since 
${\hat U}_I$ and ${\hat U}_F$ are arbitrary given unitary gates one wishes to implement, such 
a scenario will not prevail in real laboratories. 

Here we solve this problem by deriving the Euler-Poincar\'e equation of motion for the control 
Hamiltonian ${\hat H}_1(t)$, and obtain the solution in closed form. Remarkably, we find that 
the solution to the quantum Zermelo navigation problem takes the simple form: 
\begin{eqnarray}
{\hat H}_1(t) = \re^{-{\rm i}{\hat H}_0 t} {\hat H}_1(0) \re^{{\rm i}{\hat H}_0 t} , 
\label{eq:1} 
\end{eqnarray}
where ${\hat H}_1(0)$ is the initial condition such that the action generated by the total 
Hamiltonian ${\hat H}(t)={\hat H}_0+{\hat H}_1(t)$ takes ${\hat U}_I$ to ${\hat U}_F$ in 
shortest possible time. Thus, the optimal control is obtained by finding the initial direction 
$H_1(0)$ for the manoeuvre and drift along the `wind' ${\hat H}_0$. We then provide 
a scheme for finding the initial condition ${\hat H}_1(0)$. 
The results are illustrated in terms of a spin-$\frac{1}{2}$ 
system. We shall also indicate how the analysis presented here can be applied to 
situations where there are further constraints on the control Hamiltonian. 

Since the mathematical machinery required for solving the navigation problem is perhaps not 
widely accessible to the broader physics community, we begin with a brief discussion of the 
background concepts before proceeding to derive (\ref{eq:1}). To address such navigation 
problems in the calculus of variation, it is often the case that one requires the notion of a 
distance that depends not only on the location but also on the direction---a concept that goes 
outside of the realm of Riemannian geometry. Specifically, for a given curve $x^i(t)$ on the 
configuration space ${\mathfrak M}$, equipped with  a Riemannian metric, we consider the integral 
of the form 
\begin{eqnarray}
T = \int_{t_0}^{t_1} F(x^i, {\dot x}^i) \rd t 
\label{eq:2}
\end{eqnarray}
for some positive function $F$, which is assumed to be homogeneous of first degree in 
${\dot x}^i$, $F(x^i, \lambda{\dot x}^i) =\lambda F(x^i, {\dot x}^i)$ for any $\lambda>0$, so 
that $T$ is independent of the choice of the parameter $t$ along $x^i(t)$. 
Thus, $F(x^i, {\dot x}^i)$ defines, for each fixed point $x^i\in{\mathfrak M}$, a distance on the 
tangent space of ${\mathfrak M}$. In particular, the level surface $F(x, {\dot x})=1$ on 
the tangent space of 
${\mathfrak M}$ at $x$ defines the indicatrix \cite{caratheodory}. 
Now for a fixed $x$ and an arbitrary point 
$\xi$ on the tangent space of ${\mathfrak M}$ at $x$, the ray $\overrightarrow{x\xi}$ 
clearly intersects the indicatrix at a point $\rho_\xi$. Thus, conversely, for each point 
$\xi$ if we define a function $F$ according to $F(\xi) = |\xi|/|\rho_\xi|$, 
where $|\cdot|$ denotes the Euclidean norm, then we can introduce a metric, known as the 
Minkowski metric \cite{busemann},  as follows: For $\xi,\xi'$ on the tangent space 
of ${\mathfrak M}$ at $x$ the distance 
between these points is defined by $D(\xi,\xi') = F(\xi-\xi')$. In particular, the metric 
tensor defined on ${\mathfrak M}$ induced by the distance function $D$ associated with the 
fundamental function $F$ can be expressed in the form: 
\begin{eqnarray}
g_{ij}(x,{\dot x}) = \frac{1}{2} \frac{\partial^2}{\partial {\dot x}^i \partial {\dot x}^j} 
F^2(x^i, {\dot x}^i) ,
\label{eq:3} 
\end{eqnarray}
and we have $F^2 = g_{ij}{\dot x}^i{\dot x}^j$. 

In classical mechanics, often the fundamental function takes the form of the kinetic 
energy: $F^2=\gamma_{ij}(x){\dot x}^i {\dot x}^j$. Thus, 
the resulting metric $g_{ij}(x,{\dot x})=\gamma_{ij}(x)$ is independent of the direction ${\dot x}$, 
i.e. it defines a Riemannian metric on ${\mathfrak M}$, since the indicatrix is just a sphere. 
In many problems with engineering applications, such as a navigation problem, however, 
the relevant function takes a different form, and as such one is required to go beyond 
the techniques of Riemannian geometry. Realising this, Carath\'eodory suggested to his then 
PhD student Finsler to investigate the geometry of spaces equipped with such 
direction-dependent metrics \cite{finsler}. Subsequently, spaces endowed with locally 
Minkowski metrics were referred to as Finsler spaces \cite{rund}. 

Let us now turn to the classical Zermelo navigation problem of reaching a target on a manifold 
${\mathfrak M}$ equipped with a Riemannian metric $h_{ij}$ in the shortest possible time, 
in the presence of background wind $w^i$. The analysis of the problem simplifies if we 
observe that it suffices to find the locally optimal solution on the tangent space \cite{shen}. 
Specifically, for any vector $\vec\xi$ on the tangent space to ${\mathfrak M}$ at $x$ we 
can regard $|\vec\xi|_h=\sqrt{h_{ij}\xi^i\xi^j}$ as representing the time it takes to reach the 
endpoint of $\vec\xi$. Now suppose that in the absence of wind the time it takes to reach 
the destination $\vec{u}$ at full throttle is $1$ in a suitable unit (e.g., second), i.e. 
$|\vec{u}|_h=1$. In the presence of wind, with $|\vec{w}|_h<1$, however, after a journey of one 
second at full throttle the vessel will reach the point $\vec{v}=\vec{u}+\vec{w}$, instead of 
the destination $\vec{u}$. In other words, the unit sphere $|\vec{u}|_h=1$ has been displaced 
by the wind, but since $|\vec{w}|_h<1$ by assumption, the centre point $x$ remains in the interior of the sphere. 
Therefore, for any vector $\vec{\xi}$ on the tangent space the ray $\overrightarrow{x\xi}$ 
intersects the indicatrix at a point $\rho_\xi$; working out the Euclidean  norms of 
$\vec{\xi}$ and $\vec{\rho}_\xi$ and taking the ratio, a short calculation shows that the 
fundamental function takes the form (see also \cite{shen2,gibbons2}): 
\begin{eqnarray}
F(x, \xi) = \frac{\sqrt{\langle \vec{w},\vec{\,\xi}\rangle_h^2+|\vec{\,\xi}|_h^2(1-|\vec{w}|_h^2)} - 
\langle \vec{w},\vec{\,\xi}\rangle_h}{1-|\vec{w}|_h^2}, 
\label{eq:4} 
\end{eqnarray}
where $|\vec{\,\xi}|_h^2=h_{ij} \xi^i \xi^j$ and $\langle \vec{w},\vec{\,\xi}\rangle_h=h_{ij}w^i \xi^j$. 
Making use of (\ref{eq:3}), an explicit form of the metric on ${\mathfrak M}$ can be obtained. 
The calculation simplifies if one writes 
\begin{eqnarray}
\alpha_{ij}= \frac{h_{ij}}{1-|\vec{w}|_h^2}+\frac{w_i w_j}{(1-|\vec{w}|_h^2)^2}, \quad 
\beta_i = -\frac{w_i}{1-|\vec{w}|_h^2}, 
\end{eqnarray}
where $w_i=h_{ij}w^j$, so that we have $F=\sqrt{\alpha_{ij}\xi^i\xi^j}+\beta_i\xi^i$. 
The solution curves to the Zermelo navigation problem are then found by working out the 
geodesics of the metric. 

We remark that the metric of the type $\sqrt{\alpha_{ij}\xi^i\xi^j}+\beta_i\xi^i$ was introduced by 
Randers in the context of a unified theory of gravitation and electromagnetism \cite{randers}. 
However, Randers was unaware of the Finslerian nature of the metric, and attempted to 
interpret it in the Riemannian sense in the context of a five-dimensional Kaluza-Klein 
theory. Randers metrics are perhaps the most commonly investigated Finsler metrics in 
physical applications such as the electron microscope \cite{ingarden} and in propagation of sound 
and light rays in a moving medium \cite{gibbons2,luneburg,gibbons1}. 

The relevance of Finsler geometry to problems in quantum control has been observed in 
\cite{nielsen1,nielsen2}. In the presence of background fields, more recentl Russell \& Stepney 
\cite{stepney} proposed the technique of Shen \cite{shen} to be applied to the manifold 
${\mathfrak M}$ of special unitary matrices endowed with the bi-invariant trace norm. Specifically, 
working with the elements of the Lie algebra ${\hat\xi},{\hat\xi}'\in\mathfrak{su}(N)$ we 
have 
\begin{eqnarray}
\langle {\hat\xi}, \hat\xi' \rangle_h = {\rm tr}({\hat\xi}^\dagger \hat\xi') . 
\end{eqnarray} 
With this setup we wish to minimise the journey time (\ref{eq:2}) in the presence of `wind' 
given in $\mathfrak{su}(N)$ by $-\ri {\hat H}_0$, when ${\hat\xi}=-\ri {\hat H}(t)=
-\ri({\hat H}_0+{\hat H}_1(t))$. The fundamental function (\ref{eq:4}) in this quantum context 
thus reads
\begin{eqnarray}
F({\hat\xi}) = \ri \frac{\sqrt{[{\rm tr}({\hat\xi}{\hat H}_0)]^2+{\rm tr}({\hat\xi}^2)
(1-{\rm tr}({\hat H}_0^2))}-{\rm tr}({\hat\xi}{\hat H}_0)}{1-{\rm tr}({\hat H}_0^2)} ,
\label{eq:8}
\end{eqnarray}
which is just the Finslerian norm $\|\hat\xi\|$. 

To proceed we find it convenient to minimise the kinetic energy $\frac{1}{2}\int F^2 \rd t$ along 
the path, instead of $\int F \rd t$. It should be evident that the optimal path ${\hat\xi}(t)$ that 
minimises the latter also minimises the former. Writing $F^2({\hat\xi})=\|\hat\xi\|^2$ we have 
\begin{eqnarray}
\delta \|\hat\xi\|^2 = \left\langle \frac{\delta \|\hat\xi\|^2}{\delta\hat\xi} , 
\delta \hat\xi \right\rangle = 2 \| \hat\xi\| 
\left<\frac{\delta \|\hat\xi\|}{\delta\hat\xi}, \delta \hat\xi\right>,
\label{eq:9} 
\end{eqnarray}
where we have written, for any ${\hat\nu}\in{\mathfrak{su}}(N)$ and any $f({\hat\xi})$, 
\begin{eqnarray}
\left<\frac{\delta f(\hat\xi)}{\delta \hat\xi}, \hat\nu \right> = \left. 
\frac{\rd}{\rd \epsilon} f({\hat\xi}+\epsilon{\hat\nu}) \right|_{\epsilon=0},
\label{eq:10} 
\end{eqnarray}
and on account of (\ref{eq:8}) we have 
\begin{eqnarray}
\hspace{-0.5cm} 
\left<\frac{\delta \|\hat\xi\|}{\delta \hat\xi}, \hat\nu \right> &=& - \ri 
\frac{ {\rm tr}({\hat\nu}  {\hat H}_0) }{1-{\rm tr}({\hat H}_0^2)} 
\nonumber \\ && \hspace{-2.5cm} + \ri
\frac{ 
{\rm tr}(\hat\xi {\hat H}_0) {\rm tr}({\hat\nu} {\hat H}_0) + 
(1-   {\rm tr}({\hat H}_0^2))   {\rm tr}(\hat\xi {\hat\nu}) }{(1-{\rm tr}({\hat H}_0^2))
\sqrt{[{\rm tr}({\hat\xi}{\hat H}_0)]^2+{\rm tr}({\hat\xi}^2)
(1-{\rm tr}({\hat H}_0^2))}} . 
\label{eq:11}
\end{eqnarray}
Our aim is to solve 
\begin{eqnarray}
0 = \delta \left( \frac{1}{2} \int_0^1 \|\hat\xi \|^2 \rd t \right) = 
\int_0^1 \| \hat\xi\| \left<\frac{\delta \|\hat\xi\|}{\delta\hat\xi}, \delta \hat\xi \right> \rd t 
\label{eq:12}
\end{eqnarray}
with fixed end points of the curve on $SU(N)$. The constraints on the end points 
restricts admissible variations $\delta \hat\xi$. In particular, a standard result of 
Euler--Poincar\'e reduction \cite{ratiu} asserts that
\begin{eqnarray}
\delta \hat\xi = \dot{\hat\eta} - [\hat\xi, \hat\eta], 
\label{eq:13}
\end{eqnarray}
where $\hat{\eta}$ is an arbitrary curve in $\mathfrak{su}(N)$ with $\hat\eta(0) = \hat\eta(1) = 0$. 
Substituting (\ref{eq:13}) and (\ref{eq:11}) in (\ref{eq:12}) and rearranging terms, we are thus 
led to the relation: 
\begin{eqnarray}
0 &=& - \partial_t(\|{\hat\xi}\|) {\hat H}_0 - \|{\hat\xi}\|  [{\hat H}_0, {\hat\xi}] \nonumber \\ && 
+ \partial_t\left( \frac{\|{\hat\xi}\|}{\sqrt{\cdots}}  {\rm tr}({\hat\xi} 
{\hat H}_0)\right)  {\hat H}_0
+ \frac{\|{\hat\xi}\|}{\sqrt{\cdots}} {\rm tr}({\hat\xi} {\hat H}_0) [{\hat H}_0, {\hat\xi}] 
\nonumber \\ && 
 +  (1- {\rm tr}({\hat H}_0^2)) \, \partial_t\left(\frac{\|{\hat\xi}\|{\hat\xi}}{\sqrt{\cdots}} \right), 
\label{eq:14} 
\end{eqnarray} 
where we have written $\sqrt{\cdots}$ for the square-root term appearing in the numerator of 
(\ref{eq:8}). This result appears unduly complicated, however, if we take note of the fact that 
we are interested in the quantum navigation at full throttle, i.e. $\|{\hat\xi}\|=1$, then by taking 
the relevant time derivatives in (\ref{eq:14}) we deduce the Euler-Poincar\'e equation of the form:
$\dot{\hat\xi} + \ri [{\hat H}_0, {\hat\xi}]  - (\ri {\hat H}_0 + 
{\hat\xi}) {\rm tr}(\dot{{\hat\xi}}{\hat H}_0) / \sqrt{\cdots} = 0$. 
Substituting ${\hat\xi}=-\ri({\hat H}_0+{\hat H}_1(t))$ in here we thus obtain the relevant 
equation of motion for the control Hamiltonian ${\hat H}_1(t)$: 
$-\ri \dot{{\hat H}}_1 + [{\hat H}_0, {\hat H}_1] + {\hat H}_1 {\rm tr}({\hat H}_0 \dot{{\hat H}}_1) / 
\sqrt{\cdots} = 0$. 
If we eliminate the square-root term using (\ref{eq:8}) along with $F({\hat\xi})=\|{\hat\xi}\|
=1$, which gives us $\ri\sqrt{\cdots}=1+{\rm tr}({\hat H}_0{\hat H}_1)$, 
then we deduce that  
\begin{eqnarray}
\dot{{\hat H}}_1 +\ri [{\hat H}_0, {\hat H}_1] - \frac{  {\hat H}_1}{1 
+ {\rm tr}({\hat H}_1 {\hat H}_0)} \, {\rm tr}({\hat H}_0 \dot{{\hat H}}_1) = 0 , 
\label{eq:17}
\end{eqnarray}
where we have made use of the constraint that ${\rm tr}({\hat H}_1^2) =1$. 
Multiplying (\ref{eq:17}) with ${\hat H}_0$ and taking the trace, we thus deduce that 
${\rm tr}({\hat H}_0 \dot{{\hat H}}_1)=0$. We therefore conclude 
from \eqref{eq:17} that the quantum Zermelo--Euler--Poincar\'e equation takes the simple form: 
\begin{eqnarray}
\dot{{\hat H}}_1 + \ri [{\hat H}_0, {\hat H}_1] = 0. 
\label{eq:19}
\end{eqnarray}
This, however, is just the equation for a co-adjoint motion, so it can be solved, with the solution 
(\ref{eq:1}). 

It is interesting to observe that, after some lengthy but straightforward algebra, we are led to a 
simple and intuitive solution to the quantum navigation problem, namely, that we must pick the initial 
direction ${\hat H}_1(0)$ and let it be advected by the prevailing field ${\hat H}_0$. To hit the right target 
${\hat U}_F$ starting from the initial point ${\hat U}_I$, however, the initial direction ${\hat H}_1(0)$ 
has to be chosen appropriately. In what follows we shall derive an ordinary differential equation 
satisfied by the initial direction. 

We proceed by first solving the navigation problem in the absence of the wind: ${\hat H}_0=0$. 
In this case, the optimal control ${\hat H}_1$ is time independent, and the initial condition 
${\hat H}_1(0)$ can thus be obtained by taking the matrix logarithm of ${\hat U}_F{\hat U}_I^{-1}$. 
The idea behind our scheme is to gradually increase ${\hat H}_0$ from zero to the level specified 
by the problem, while calculating, for each increment of ${\hat H}_0$, the optimal control 
Hamiltonian that solves the Zermelo problem with that wind. Clearly, as $\hat{H}_0$ is increased, 
$\hat{H}_1(0)$ has to be adjusted as well, or else the target gate will be missed. Moreover, the 
trajectory might take slightly more or slightly less time. Hence, the duration of the trajectory needs 
also be adapted.  

With this in mind, let us calculate how the final gate varies when the wind, the initial control, and 
the terminal time are adjusted infinitesimally. Let ${\hat U}(t)$ be a curve in $SU(N)$ starting at 
${\hat U}_I$ satisfying $\partial_t{{\hat U}} = \hat{\xi} {\hat U}$ for some curve $\hat{\xi}$ in 
$\mathfrak{su}(N)$, and fix a time $s$. If $\hat{\xi}(t) + \epsilon \delta \hat{\xi}(t)$ is a variation 
of $\hat{\xi}$, then ${\hat U}(s)$ varies as
\begin{eqnarray}
\delta   {\hat U}(s) = {\hat U}(s) \int_0^s {\hat U}(t)^{-1} \delta \hat{\xi}(t) {\hat U}(t)\, 
{\rm d}t . 
\label{eq:20}
\end{eqnarray}
This follows from adapting Lemma 2.4 of \cite{BGBHR} to the present context. To proceed, 
let us write $\hat{H}_1(0, \lambda)$ for the optimal initial control and $T_\lambda$ for the 
duration of the trajectory when the wind is given by $\lambda \hat{H}_0$, $\lambda\in[0,1]$. 
Let us further denote by ${\hat U}_\lambda(t)$ the corresponding geodesic curve in $SU(N)$. 
In what follows we shall write derivatives with respect to $\lambda$ as $T'$, 
${\hat U}_\lambda'$, and 
so on. Notice that ${\hat U}_\lambda(T_\lambda)$ equals the target gate $\hat{U}_F$ for all 
$\lambda$. Hence, ${\hat U}_\lambda'(T_\lambda) = 0$. Using \eqref{eq:20}, we thus obtain 
\begin{eqnarray}  
0 &=& {\hat U}_\lambda^{-1}(T_\lambda) {\hat U}_\lambda'(T_\lambda)  \nonumber \\ 
&& \hspace{-0.8cm} =
\int_0^{T_\lambda}\!\!\! {\hat U}_\lambda(t)^{-1} \hat{\xi}_\lambda'(t) {\hat U}_\lambda(t)\, 
{\rd }t + T_\lambda' {\hat U}_\lambda^{-1}(T_\lambda) \hat{\xi}_\lambda(T_\lambda) 
{\hat U}_\lambda(T_\lambda). \nonumber \\ 
\label{eq:21} 
\end{eqnarray}
Recall that $\hat{\xi}_\lambda(t) = -\ri (\lambda \hat{H}_0 + \hat{H}_1(t, \lambda))$, where 
$\hat{H}_1(t, \lambda)$ is given by \eqref{eq:1}. Therefore, upon differentiation, 
$
  \hat{\xi}_\lambda'(t) =-\ri  \re^{-{\rm i} \hat{H}_0 \lambda t}(\hat{H}_0 + \ri t 
  [\hat{H}_1(0,\lambda), \hat{H}_0] \!+\! 
  \hat{H}_1'(0,\lambda) ) \re^{{\rm i} \hat{H}_0 \lambda t}, 
$
from which it follows that \eqref{eq:21} is a linear equation in $T_\lambda'$ and 
${\hat H}_1'(0,\lambda)$, admitting a unique solution for each $\lambda$ once 
the linear constraint ${\rm tr}({\hat H}_1(0,\lambda) {\hat H}_1'(0,\lambda))=0$ is taken 
into account. Finally, $T_\lambda'$ and ${\hat H}_1'(0,\lambda)$ can be integrated up 
to $\lambda = 1$ starting from the wind-free solution $\lambda = 0$. The optimal 
initial control is then given by ${\hat H}_1(0, 1)$, and the trajectory 
is traversed in time $T_1$. 

\begin{figure}[t!]
\begin{overpic}[scale=0.18]{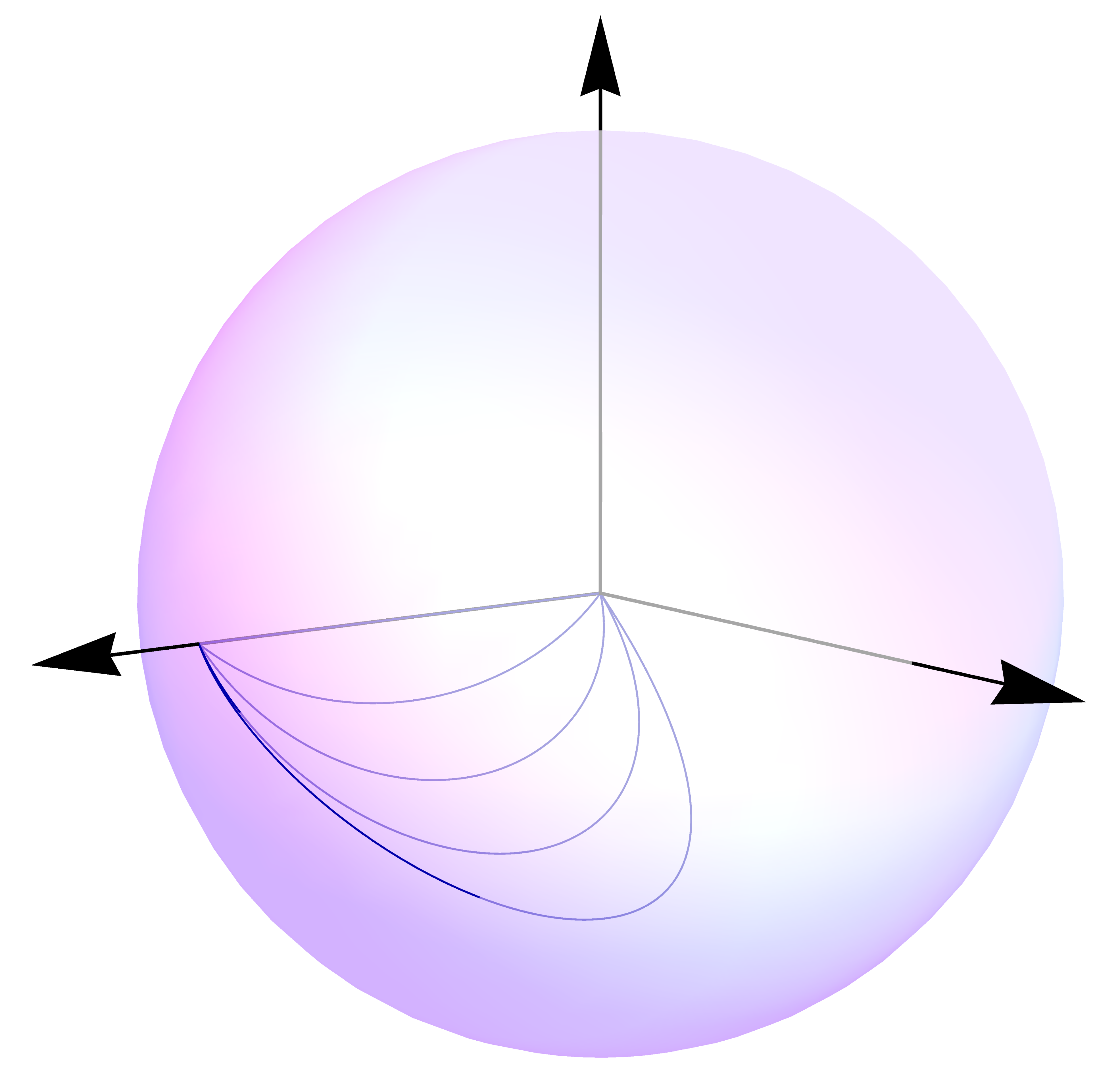}
\put(54.5, 44.5){$\hat{U}_I$}
\put(52.5, 42.3) {$\bullet$}
\put(16, 41) {$\hat{U}_F$}
\put(17, 37.7) {$\bullet$}
\put(6, 32.5) {$x$}
\put(49, 90) {$z$}
\put(95.5, 37) {$y$}
\end{overpic}
\footnotesize
\caption{\label{fig:1}
\emph{Optimal generation of target unitary gate}. 
The time-optimal trajectories ${\hat U}(t)$ are shown for various wind-strengths $\omega = 
0, 0.25, 0.5, 0.75, 1$, as  curves in the rotation group using the standard covering map. 
The centre of the sphere corresponds to the initial gate ${\hat U}_I={\mathds 1}$, while the 
terminal point that lies on the surface of the sphere is the target gate ${\hat U}_F= -
\ri{\hat\sigma}_x$. The direction of the vector joining the centre ${\mathds 1}$ to a point 
${\hat U}(t)$ on a given curve represents the axis of rotation, whereas the radius of the 
vector represents the angle of rotation. The sphere upon which the target gate lies thus 
has radius $\pi$. 
}
\end{figure}
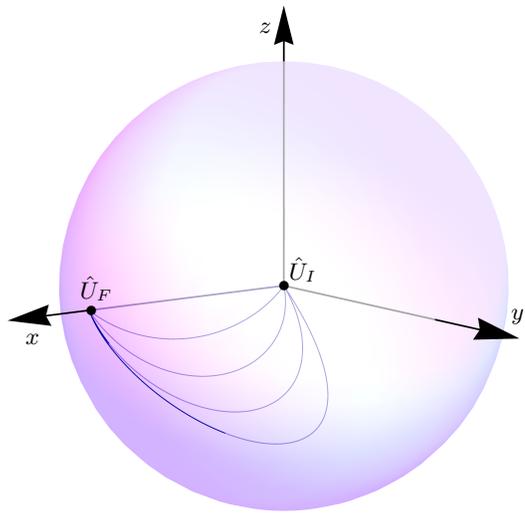

In summary, we have derived the Euler-Poincar\'e equation (\ref{eq:19}) associated 
with the quantum Zermelo navigation problem introduced in \cite{stepney}. The equation 
of motion is surprisingly simple, and admits an elementary solution (\ref{eq:1}). 
We have provided a scheme which allows for the determination of the initial control 
Hamiltonian ${\hat H}_1(0)$ required to hit the correct target point ${\hat U}_F$. On 
account of linearity, our scheme can easily be implemented in practice. With the solution 
(\ref{eq:1}) at hand, optimal quantum control with finite energy resources becomes 
feasible under the presence of external field or potential that might be difficult to eliminate 
in laboratories. 
As an illustrative example let us consider the control of a spin-$\frac{1}{2}$ 
system, where the objective is to transform ${\hat U}_I={\mathds 1}$ into $\hat{U}_F = 
\re^{-{\rm i} \pi {\hat\sigma}_x / 2} = -\ri {\hat\sigma}_x$, under the influence of an external field 
$\hat{H}_0 = -\omega {\hat\sigma}_z$, where ${\hat\sigma}_x, {\hat\sigma}_y, {\hat\sigma}_z$ 
are the Pauli matrices. In this example a closed-form expression for the optimal initial 
Hamiltonian ${\hat H}_1(0)$ can be obtained on account of the relation (cf. \cite{stepney2,brody5}) 
$\hat{U}_F = -\ri {\hat\sigma}_x = \re^{{\rm i} \omega {\hat\sigma}_z T} \re^{-{\rm i} \hat{H}_1(0) T}$, 
which follows from (\ref{eq:1}). 
Specifically, a short calculation shows that $\hat{H}_1(0) = \frac{1}{\sqrt{2}} {\boldsymbol n}\cdot
\hat{\boldsymbol{\sigma}}$ and $T=\pi/\sqrt{2}$, where the unit vector ${\boldsymbol n}$ is given 
by ${\boldsymbol n}=(\cos(\omega T), \sin(\omega T),0)$. The resulting unitary orbit 
${\hat U}(t)$ is sketched in figure~\ref{fig:1} for a range of values of $\omega$.  

We conclude by remarking that in the presence of additional constraints on the control Hamiltonian 
that limit the implementability of \eqref{eq:1}, it suffices to include them in the maximisation of $F^2$ 
by use of Lagrange multipliers. It then follows that the solution \eqref{eq:1} remains valid, except 
that the initial control $H_1(0)$ is replaced by a time-dependent one (cf. \cite{brody5}). More 
precisely, what the solution \eqref{eq:1} shows is that it is possible to switch to a frame that moves in 
the counter direction to the wind  so that the analysis of constrained optimisation 
performed, for example in \cite{hosoya}, with time-dependent constraints, becomes applicable. 
In this manner the 
solution to the Zermelo navigation problem presented here can be extended straightforwardly 
to accommodate 
further constraints that one might encounter for instance in systems involving a large number of 
coupled spins where controllable degrees of freedom are typically limited.

\vspace{0.2cm} 

\begin{acknowledgments}
We thank Gary Gibbons for drawing our attention to \cite{stepney,gibbons2,gibbons1}. 

Note added: 
Russell \& Stepney have independently obtained the solution (\ref{eq:1}) to the quantum Zermelo 
navigation problem \cite{stepney2}, using a theorem of \cite{robles} on geodesics of Randers 
spaces (rather 
than deriving and solving the Euler--Poincar\'e equation as we have done here). 
\end{acknowledgments}

\end{document}